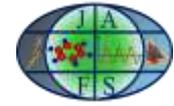

# VARIATIONAL STUDY OF S-SHELL LAMBDA HYPERNUCLEAR SYSTEM

Bhupali Sharma*

*Dept. of Physics, Arya Vidyapeeth College, Guwahati-781016*
*For correspondence. (bhupalisharma@gmail.com)

Abstract:: Theoretical study on hypernuclear systems is important to know the nature of hyperon-nucleon and hyperon-hyperon interaction as only hypernuclear systems give the scope of knowing these interactions. A hypernucleus, in addition to the nucleons contains at least one hyperon which is a strange particle composed of quarks. A hypernucleus is produced mostly in heavy ion collisions and it undergoes weak decay. Experimental detection of hypernuclear events are rare and this makes the study of hypernuclear physics more challenging. Hypernuclear physics has a close association with astrophysics as hyperon-nucleon and hyperon-hyperon interactions are found to play important role in the interiors of neutron stars. The core of neutron stars contains strange quark matter and therefore, study of hyperon involved potentials are essential for the determination of the composition of neutron star matter. But, there is a scarcity of data from hyperon-nucleon scattering experiments. Also, since it is impossile to have hyperon-hyperon scattering experiments, the direct determination of the baryon-baryon interaction strength is extremely difficult. Therefore theoretical models play important role in unfolding the mysteries of hyperon-nucleon and hyperon-hyperon interaction. In this study, binding energies of hypernuclear systems calculated using different two-body lambda-nucleon and three-body lambda-nucleon-nucleon interactions have been analysed. Also effect of lambda-lambda potential on the binding energy of hypernuclear system have been analysed. In this few-body study, we have employed Variational Monte Carlo technique for calculation of the binding energies of different hyperclear systems.

Keywords: hypernuclear system; few-body study; Variational Monte Carlo technique

PACS: 21.80.+a, 21.60.GX, 21.45.-v, 21.30.-x

## 1. Introduction:

A hypernucleus, in addition to the nucleons contains at least one hyperon which is a strange particle. Hypernucleus is produced mostly in heavy ion collisions and it undergoes weak decay. In general, a single and a double lambda hypernucleus is represented by, $_\Lambda^A Z$ and $_{\Lambda\Lambda}^A Z$ where 'Z' and 'A' denote respectively the atomic number and mass number of the parent atom. A lambda hypernucleus has one or more lambda particle occupying the same quantum state occupied already by the nucleons. The strange hyperon 'Λ' has a lifetime of $10^{-10}$ s and therefore a hypernucleus decays very rapidly. This makes experimental detection of hypernuclear events very rare and therefore the study of hypernuclear physics is very challenging. Study of hyperceus is important mainly for the following reasons, to understand the nature of hyperon-nucleon and hyperon-hyperon interactions, to understand baryon-baryon interaction in general, enriching our knowledge about role of strangeness in a nuclear medium of different densities in physical observables starting from deuteron to neutron stars etc. Study of hyperon involved potentials are essential for the determination of the composition and properties of neutron star matter (eg. Cooling rate of neutron stars, mass-radii relations etc.), as hyperons are present in the cores of neutron stars. Since the first report of hypernuclear event by M. Danysz and J. Pniewski[1] in 1953, lots of theoretical as well as experimental work has been done.

Earlier we have performed variational Monte-Carlo studies on different hypernuclear systems with different potential models to study the effect of parameters of the potential models on the binding energy of the hypernuclear systems [2,3,4]. In the present study we revisit the hypernuclear system $_{\Lambda\Lambda}^4 H$ with the potential model ΛN4 of ref[8]. We discuss the role of ΛNN interaction parameters on the stability of the double hypernuclear system $_{\Lambda\Lambda}^4 H$ From the study we find that the stability of the hypernuclear system depends on the three-body ΛNN potential crucially.

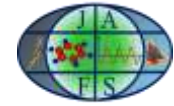

## 2. Potentials and wavefunction:

*NN and NNN potential :*

For the nuclear part of the Hamiltonian, we use *Argonne V₁₈* NN[6] and *Urbana IX* NNN[7] potentials .

*ΛN  and ΛNN potential :*

We use phenomenological potential consisting of central, Majorana space-exchange and spin-spin ΛN components for ΛN potential[5], and  it is given by,

$$V_{\Lambda N} = (V_c(r) - \bar{V} T_\pi^2(r))(1 - \varepsilon + \varepsilon P_x) + \frac{1}{4} V_\sigma T_\pi^2(r) \sigma_\Lambda . \sigma_N \qquad (1)$$

Here $P_x$ is the majorana space-exchange operator and $\varepsilon$ is the space exchange parameter which is taken as 0.2[8]. $V_c(r)$,  $\bar{V}$ and $V_\sigma$  are respectively Wood-saxon core, spin-average and spin-dependent strength and $T_\pi^2(r)$ is one-pion tensor shape factor.

In the  ΛNN potential, there are two terms, a two-pion exchange part and a dispersive part[5]. The two-pion exchange part of the interaction is given by

$$W_P = -\frac{1}{6} C_P (\tau_i . \tau_j) \{ X_{i\Lambda} . X_{j\Lambda} \} Y_\pi(r_{i\Lambda}) Y_\pi(r_{j\Lambda}) \qquad (2)$$

Here $X_{k\Lambda}$ is the one-pion exchange operator given by,

$$X_{k\Lambda} = (\sigma_k . \sigma_\Lambda) + S_{k\Lambda}(r_{k\Lambda}) T_\pi(r_{k\Lambda})$$

with

$$S_{k\Lambda} = \frac{3(\sigma_k . r_{k\Lambda})(\sigma_\Lambda . r_{k\Lambda})}{r_{k\Lambda}^2} - (\sigma_k . \sigma_\Lambda)$$

The dispersive part of the ΛNN potential is given by,

$$V_{\Lambda NN}^{DS} = W_0 T_\pi^2(r_{i\Lambda}) T_\pi^2(r_{j\Lambda})[1 + \frac{1}{6} \sigma_\Lambda . (\sigma_i . \sigma_j)] \qquad (3)$$

Here $Y_\pi(r_{k\Lambda})$ and $T_\pi(r_{k\Lambda})$ are the usual Yukawa and tensor functions with pion mass, $\mu = 0.7$ fm$^{-1}$ ; $C_p$ and  $W_0$ are ΛNN interaction parameters.

We revisit the hypernuclear system $_{\Lambda\Lambda}^4 H$ using the  potential model given in Table 1.The values of the  ΛN and ΛNN interaction parameters, viz. ε, $C_p$ & $W_0$   selected with the criterion of giving bound state for     $_\Lambda^3 H$. For the potential model ΛN4, the spin-average and spin-dependent strength of the ΛN potential are kept same with spin-average strength  $\bar{V}$ = 6.150 Mev and spin-dependent strength $V_\sigma$ = 0.176 Mev, same as in ΛN1[2,3]. The values of the interaction parameter of our earlier potential model  *ΛN1*[2,3]  are listed in Table 2. The ΛN and ΛNN potential parameters for our preferred model denoted by  ΛN4 [8]are listed in Table1. $C_p$ and $W_0$ are the strength parameters of the two-pion and dispersive parts of the ΛNN potential. The values of the three body ΛNN interaction parameters have been reduced in this preferred potential model  ΛN4.

Table 1:  ΛN and ΛNN interaction parameters for ΛN4. Except for ε, all other   quantities are in  MeV.

| ΛN | $\bar{V}$ | $V_\sigma$ | ε | $C_p$ | $W_0$ |
|---|---|---|---|---|---|
| ΛN4 | 6.150 | 0.176 | 0.2 | 0.7 | 0.012 |

Table 2:  ΛN and ΛNN interaction parameters for ΛN1. Except for ε, all other   quantities are in  MeV.

| ΛN | $\bar{V}$ | $V_\sigma$ | ε | $C_p$ | $W_0$ |
|---|---|---|---|---|---|
| ΛN1 | 6.150 | 0.176 | 0.2 | 0.15 | 0.028 |

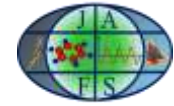

*ΛΛ potential :*

For ΛΛ potential, we use low-energy phase equivalent Nijmegen interactions represented by a sum of the three Gaussians[9,10,11],

$$V_{\Lambda\Lambda} = v^{(1)} \exp(-r^2 / \beta_{(1)}^2) + v^{(2)} \exp(-r^2 / \beta_{(1)}^2) + v^{(3)} \exp(-r^2 / \beta_{(3)}^2)$$

Here the strength parameters $v^i$ and the range parameters $\beta_i$ are,

| Sl. No. | $\beta_i$ (fm) | $v^i$ (Mev) |
|---------|---------------|-------------|
| 1 | 1.342 | -21.49 |
| 2 | 0.777 | -379.10 |
| 3 | 0.350 | 9324.00 |

The variational wave function is represented by,

$$|\Psi_v\rangle = \left[ 1 + \sum_{i<j<k} (U_{ijk} + U_{ijk}^{TNI}) + \sum_{i<j,\Lambda} U_{ij,\Lambda} + \sum_{i<j} U_{ij}^{LS} \right] \prod_{i<j<k} f_c^{ijk} |\Psi_p\rangle \qquad (4)$$

Here, $|\Psi_p\rangle$ is the pair wave function[2,3] given by

$$|\Psi_p\rangle = S \prod_{i<j} (1 + U_{ij}) S \prod_{i<\Lambda} (1 + U_{i\Lambda}) |\Psi_J\rangle \qquad (5)$$

The Jastrow wave function for lambda hypernuclei is given by,

$$|\Psi_J\rangle = [ \prod_{i<j<k} f_c^{ijk} \prod_{i<\Lambda} f_c^{i\Lambda} \prod_{i<j} f_c^{ij} ] |\Psi_{JT}\rangle \; |\varphi\rangle$$

Here f' s are the central correlation functions and |φ> is an antisymmetric wave function of the lambda particle. $|\Psi_{JT}\rangle$ is the spin and isospin wavefuntion of the s-shell nucleus.

3. Variational Monte Carlo technique:

Variational Monte Carlo method is used to find the ground state energy and binding energy of different hypernuclear systems. This technique is based on variational principle and is used to find ground state energy of a system by varying the different parameters variationally.

Variational principle states that, the approximate value of a Hamiltonian, calculated using trial wave-function is never lower in value than the true ground state energy

$$E = \frac{\langle \Psi | H | \Psi \rangle}{\langle \Psi | \Psi \rangle} \geq E_o \qquad (7)$$

where $E_0$ is the true ground state energy of the system.

To find the true ground state energy , a suitably parametrized trial wave function is selected which is a function of position, spin, isospin and other intrinsic variables and parameters. This trial wave function is used to find the upper bound to the energy using metropolis algorithm. The minimum energy is searched by calculating energy differences for wave functions using configurations generated by random walk. Energy expectation values are calculated by varying variational parameters one or two at a time. The minimum energy so obtained is taken as the true ground state energy of the system.

The binding energy ($B_\Lambda$ ) formulae for single and double hypernuclear system are given by,

$$- B_\Lambda (_\Lambda^A Z) = E(_\Lambda^A Z) - E(^{A-1} Z) \qquad (8)$$

$$- B_{\Lambda\Lambda} (_{\Lambda\Lambda}^A Z) = E(_{\Lambda\Lambda}^A Z) - E(^{A-2} Z) \qquad (9)$$

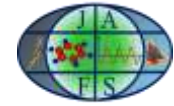

## 4. Results and discussion:
·

The binding energy results for the hypernuclear systems $_{\Lambda\Lambda}^{4}H$ with the potential model $\Lambda$N4 are tabulated in Table 3. We have also presented the results for $_{\Lambda}^{3}H$ the two potentials. We have compared the present result with experimental value and our earlier result on $_{\Lambda\Lambda}^{4}H$ with the potential model $\Lambda$N1 [Ref. 3].

Table3: Binding energy ( $B_{\Lambda}$ and $B_{\Lambda\Lambda}$ ) results for $_{\Lambda}^{3}H$ and $_{\Lambda\Lambda}^{4}H$ . All quantities are in MeV.

| Potential | Term | $_{\Lambda}^{3}H$ | $_{\Lambda\Lambda}^{4}H$ |
|---|---|---|---|
| $\Lambda$N4 | E | -2.39(01) | -2.30(01) |
| | $V_{\Lambda N}$ | -2.86(09) | -2.12(04) |
| | $V_{\Lambda NN}$ | -0.15(01) | -0.07(01) |
| | SEC | 0.05(00) | 0.07(00) |
| | $B_{\Lambda}$ | 0.17(01) | |
| | $B_{\Lambda\Lambda}$ | | 0.08(01) |
| | | | |
| *Experimental* | | *0.13* | |
| *$\Lambda$N1 [Ref. 3]* | | *0.34(01)[3]* | *0.38(03)* |

The potential model $\Lambda$N4 contains both space exchange part of $\Lambda$N potential and non zero values of the parameters $C_p$ & $W_0$ of $\Lambda$NN potential but with reduced values. The binding energy for $_{\Lambda\Lambda}^{4}H$ is found to be very small, 0.08(01) with this potential model. Therefore higher values of these two parameres give more stable $_{\Lambda\Lambda}^{4}H$.

## 5. Conclusions:

The results show that the decreased values of $\Lambda$NN interaction parameters reduces the calculated binding energy. Thus, in addition to the two-body $\Lambda$N interaction parameters and $\Lambda\Lambda$ interaction parameters, the three-body $\Lambda$NN interaction parameters are important for bound state of $_{\Lambda\Lambda}^{4}H$ .

Acknowledgements:

Bhupali Sharma acknowledges the facilities at Department of Physics, Arya Vidyapeeth College, Guwahati for carrying out the work.

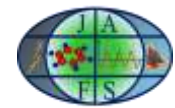